\documentclass[%
 reprint,
 amsmath,amssymb,
 aps,
 prl,
 superscriptaddress
]{revtex4-2}

\usepackage{graphicx}
\usepackage{dcolumn}
\usepackage{bm}
\usepackage{xr} 
\usepackage[utf8]{inputenc}

\begin{document}

\title{
Second-order Phase Transition in Phytoplankton Trait Dynamics
}

\author{Jenny Held}
\affiliation{Eawag, Swiss Federal Institute of Aquatic Science and Technology, \"Uberlandstrasse 133, 8600 D\"ubendorf, Switzerland}
 
\author{Tom Lorimer}
\affiliation{Eawag, Swiss Federal Institute of Aquatic Science and Technology, \"Uberlandstrasse 133, 8600 D\"ubendorf, Switzerland}

\author{Francesco Pomati}
\affiliation{Eawag, Swiss Federal Institute of Aquatic Science and Technology, \"Uberlandstrasse 133, 8600 D\"ubendorf, Switzerland}

\author{Ruedi Stoop}
\affiliation{Institute of Neuroinformatics, Winterthurerstrasse 190, 8057 Zurich, Switzerland}
 
\author{Carlo Albert}\email{carlo.albert@eawag.ch}
\affiliation{Eawag, Swiss Federal Institute of Aquatic Science and Technology, \"Uberlandstrasse 133, 8600 D\"ubendorf, Switzerland}

\date{\today}

\begin{abstract}
Key traits of unicellular species, like cell size, often follow scale-free or self-similar distributions, hinting at the possibility of an underlying critical process. However, linking such empirical scaling laws to the critical regime of realistic individual-based model classes is difficult.
Here we reveal new empirical scaling evidence associated with a transition in the population and chlorophyll dynamics of phytoplankton. 
We offer a possible explanation for these observations by deriving scaling laws in the vicinity of the critical point of a new universality class of non-local cell growth and division models. This `criticality hypothesis' can be tested through new scaling predictions derived for our model class, for the response of chlorophyll distributions to perturbations.
The derived scaling laws may also be generalized to other cellular traits and environmental drivers relevant to phytoplankton ecology.

\end{abstract}

\maketitle

\textbf{
Phytoplankton is responsible for up to half of all primary production on Earth, thus playing an essential role for the global nutrient cycles. 
We propose that essential aspects of the phytoplankton response to changing environmental conditions can be described, in a robust quantitative manner, by scaling laws.
While distributions of energetically relevant traits, such as cell size, are known to exhibit scaling properties within natural communities and across different species, we present here first experimental evidence that the response of cellular chlorophyll content to changing light conditions has two scaling regimes.
We argue that the transition between these regimes can be seen as a phase transition near the critical point of a generic class of growth-division models.
Our criticality hypotheses also allows for further testable scaling predictions and generalizations to other environmental and cellular variables.
}

Empirical scaling is routinely observed in complex natural and artificial systems \citep{banavar2007,Chialvo2010,lux1999scaling,faloutsos1999power}.
It often manifests itself in the form of power-law distributions \cite{rinaldo_2002_MicrobialSizeSpectra, maranon_2015_CellSizePhytoMetabolism, CavenderBares_2001_MicrobialSizeSpectra} or self-similar families of distributions that collapse onto a universal curve upon re-scaling \cite{hosoda_2011_GrowthDivisionDataCollapse, giometto2013scaling, salman_2012_ProteinDataCollapse, IyerBiswas_2014_SizeDataCollapseTemperature}. 
This ubiquity of scale-invariance in systems that are far from equilibrium has been explained by {\em self-organized criticality} \cite{bak1987self}, or {\em generic scale-invariance} \cite{grinstein_1991_genericScaleInvariance} - 
two concepts that lead to dynamically diverging correlation lengths. Biological systems seem to exploit the ensuing diverging susceptibility to external stimuli 
\citep{kern2003essential,stoop2016auditory,kinouchi_2006_excitableNetwork}.
The diverging correlation lengths also render the scaling laws independent of microscopic system details, which allows us to explain them by means of relatively simple mechanistic model classes. 
The specific model classes, in turn, provide us with a theoretical basis for further testable scaling predictions and robust generalisations of the observed scaling laws.
For many scaling laws observed in natural complex systems however, such model classes are still unknown.



In our contribution we analyze observed scaling in cellular properties of phytoplankton, such as size or chlorophyll content. Phytoplankton are responsible for up to half of all primary production on Earth \cite{field1998primary}, rendering these cellular ’traits’ of particular importance to global nutrient cycles. 
In natural communities, phytoplankton exhibit size-scaling that holds over almost 3 orders of magnitude \citep{rinaldo_2002_MicrobialSizeSpectra, maranon_2015_CellSizePhytoMetabolism, CavenderBares_2001_MicrobialSizeSpectra} and in experiments, individual species’ size distributions were found to be self-similar \citep{giometto2013scaling}. 
Here we provide new experimental evidence of scaling in cellular chlorophyll-a content of two different types of phytoplankton, where we observe two qualitatively distinct ’phases’ of chlorophyll-a scaling and population growth. 
We explain these observations with a class of generic growth-division models that exhibits a critical point associated with a second-order phase transition. The scaling laws in the vicinity of this critical point also provide an explanation for the aforementioned self-similarity of size distributions \citep{giometto2013scaling} and provide new, testable predictions of scaling in perturbation responses.

\begin{figure}[h!]
    \centering
    \includegraphics[width=1\columnwidth]{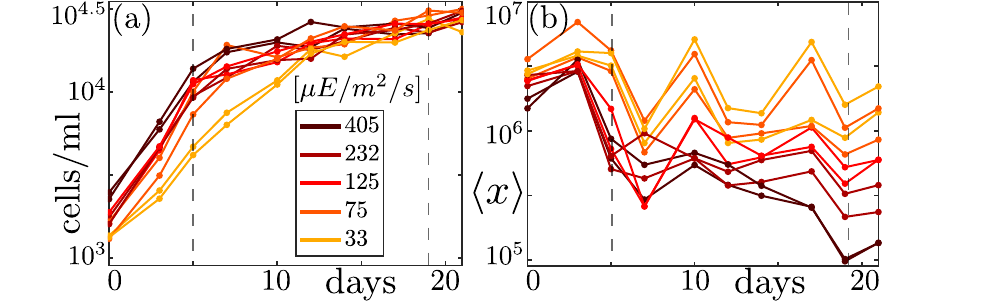}
    \vspace{-4mm}
    \caption{Dynamics of cell density (a) and mean cellular chlorophyll-a (b), for different light settings measured in photon irradiance. A transition from fast to slow exponential population growth accompanied by a decrease in mean chlorophyll-a content is observed. Dashed lines indicate days 5 and 19 characteristic for the two phases.}\label{fig:timeDev}
\end{figure}

Our experimental data involves two species of phytoplankton, {\em Kirchneriella subcapitata} (green algae) and {\em Microcystis aeruginosa} (cyanobacteria), grown in monocultures under 14h-10h light-dark cycles with six different light intensities, and a fixed initial nutrient concentration that was not replenished (cf. the previous work of Ref. \cite{fontana_2019_lightExp}). 
We focus here on {\em Kirchneriella subcapitata} - similar results also hold for {\em Microcystis aeruginosa} (see Appendix). In the experiment, cultures were stirred at regular intervals and samples were passed through a scanning flow cytometer, providing estimates of cell number density and cellular chlorophyll-a content. 
We discarded measurements corresponding to non-living particles \cite{thomas2018}, as 
well as outliers (see Appendix and discussion below).
For our analysis, the cytometer’s 677–700nm fluorescence curve integrals characteristic for the chlorophyll-a content were power-law transformed \cite{Alvarez_2017_FluorescenceSizeScaling} (which leaves scaling properties invariant). We focus on the 5 highest light intensities.

\begin{figure}[h!]
    \centering
    \includegraphics[width=1\columnwidth]{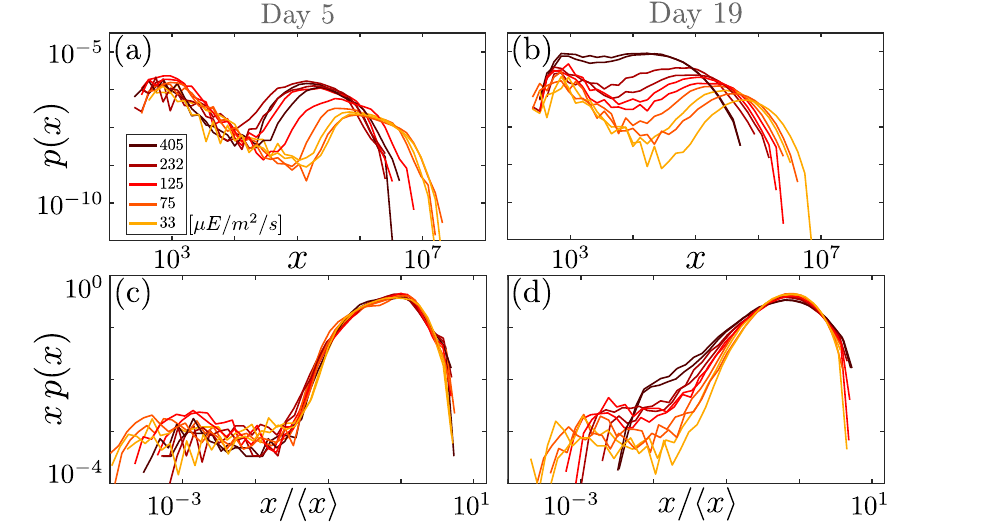}
    \vspace{-3mm}
    \caption{
    Chlorophyll-a distributions (a,b) and their rescaled versions (c,d) at days 5 (a,c) and 19 (b,d), respectively, show that the data collapse during the fast growth phase (c) breaks down in the slow growth phase (d).
    }\label{fig:distrColl}
\end{figure}

\begin{figure}[h!]
    \centering
    \includegraphics[width=1\columnwidth]{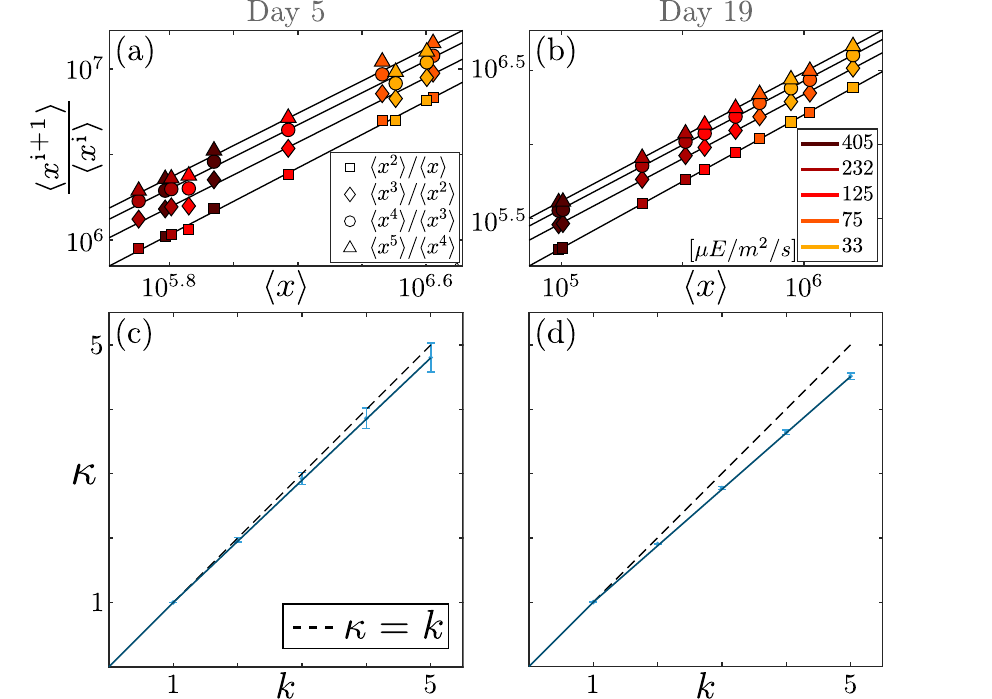}
    \vspace{-4mm}
    \caption{
    Scaling of successive moment ratios (a,b), and derived moment scaling exponents from $\langle x^k\rangle\sim\langle x\rangle^{\kappa(k)}$ (c,d), at days 5 and 19, respectively, showing consistency with trivial (c) and anomalous (d) moment scaling (error bars indicate 95$\%$ confidence intervals of corresponding moment ratio fits).}\label{fig:momentScalingPlusRatios}
\end{figure}

The observed population growth (Fig.~\ref{fig:timeDev} (a)) suggests the existence of two distinct growth phases, characterized by fast and slow exponential population growth rates, respectively. 
The transition between these phases, which occurs around day 7 (the precise timing varies between cell cultures) may be induced by nutrient depletion. 
Towards the fully developed fast growth phase (day 5), the family of chlorophyll distributions corresponding to the different light intensities shows a neat data collapse (Fig.~\ref{fig:distrColl}, left panels): A single scale, given by the mean chlorophyll content $\langle x \rangle_i$, suffices to characterize the $i$th distribution of the family. 
Because of this, the higher moments scale trivially with the mean chlorophyll content as $\langle x^k\rangle_i\sim\langle x\rangle_i^k$, which we verified in our data to sufficient accuracy up to the fifth moment (Fig.~\ref{fig:momentScalingPlusRatios}, left panels). 
In contrast, chlorophyll distributions from the slow growth phase cannot be brought to collapse upon re-scaling with a single scale (Fig.~\ref{fig:distrColl}, right panels, for the fully developed slow growth phase). Here, the higher moments scale as $\langle x^k\rangle_i\sim\langle x\rangle_i^{1+\gamma (k-1)}$, with $\gamma$ significantly smaller than one (Fig.~\ref{fig:momentScalingPlusRatios}, right panels). 

To arrive at a model for these observations, because of the exponential population growth, we neglect cell interactions and consider only cell growth, reproduction and death.
Each cell in the population is completely described by a single positive value $x$ that indicates any continuously varying trait at the cellular level like biomass or chlorophyll content (cell division, where a cell with trait value $x$ splits into two daughter cells of trait value $x/2$ each, exempted).
The rates of the different processes are assumed to be functions of $x$, which of course depend on the species and on the environmental conditions. The growth rate of an individual cell, $g(x)$, carries the dimension of trait per time, whereas cell division and cell death rates, $b(x)$ and $d(x)$, respectively, are frequencies. 
The division process renders this class of models in trait space inherently {\em non-local}. 
Under the assumption of a large number of cells, we may neglect cell-number fluctuations, and describe the system by a deterministic trait- and time-dependent cell number density, $\psi(x,t)$, that satisfies the linear, non-local mean-field equation \citep{fredrickson_1967_GrowthDivisionPDE, sinko_1967_GrowthDivisionPDE, sinko_1971_GrowthDivisionPDE, hall1989functional,hall1990functional},
\begin{multline}\label{eq:HallWake}
\frac{\partial}{\partial t}\psi(x,t) =
-\frac{\partial}{\partial x}(g(x)\psi(x,t))
+4b(2 x)\psi(2x,t)
\\
-(b(x)+d(x))\psi(x,t)
\,.
\end{multline}{}
The factor $4$ in the second term on the r.h.s. of Eq.~(\ref{eq:HallWake}) is the result of a factor $2$ due to the cells splitting into two and of a factor $2$ from the Jacobian of the transformation $x\mapsto 2x$.
Eq. (\ref{eq:HallWake}) has been used to explain the data collapse of size distributions of different species \cite{hosoda_2011_GrowthDivisionDataCollapse}, as well as the exponential tails of cellular protein distributions \cite{friedlander_2008_GrowthDivisionExpTail}.
These approaches only considered {\em one} trait scale. 
Because our observed scaling laws suggest more than one relevant trait scale, we introduce {\em two} scales associated with processes limiting too large and too low trait values. For simplicity, we omit the death process (for now) and suppress too large trait values by an increased division rate (for large $x$) and too small trait values by an increased growth rate (for small $x$):
\begin{equation}\label{eq:SI_ModelA}
   g(x)=\omega_1x(1+ \tilde g(u/x))\,,\quad b(x)=\omega_2(1+\tilde b(x/v))\,.
\end{equation}
Here, parameters $u$ and $v$ define the lower and upper limiting trait scales, respectively, and $\tilde g$ and $\tilde b$ are monotonously increasing analytic functions that can be expanded into Taylor series $\tilde g(y)=y+\mathcal{O}(y^2)$ and $\tilde b(y)=y+\mathcal{O}(y^2)$ about the origin.
While the dependence of growth and division rates of phytoplankton cells on size (or chlorophyll content) is not known, evidence from bacteria suggests that cells grow exponentially \citep{OsellaEColiDivision,WangEColiGrowth} and that the division rate increases with increasing size \citep{KennardEColiDivision}, which is compatible with Eqs.~(\ref{eq:SI_ModelA}), for $x\gg u$.

We describe the time and trait scales of the trait dynamics by the two dimensionless parameters $\tau =\omega_2/\omega_1-1$ and $h=u/v$, respectively. Henceforth, we work in units where $\omega_1=1$ and $v=1$ unless otherwise specified.
The dynamics are approximately trait-scale invariant where $h\ll x\ll 1$, so in this region the equilibrium trait distribution approximately follows a power law. After a time that is sufficiently long compared to $|\tau|^{-1}$ (see Eq.~(\ref{eq:SI_slowingdown}) below), the cell-number density is thus described by the ansatz
\begin{equation}\label{eq:FSS}
    \psi(x,t)=e^{r(\tau,h) t}x^{-\alpha(\tau)}\mathcal{F}(h/x,x)\,,
\end{equation}
where $r$ is the effective population growth rate and $\mathcal{F}$ is a cut-off function returning approximately a constant unless one of its arguments approaches unity (i.e., $x$ approaches the limits set by $u$ and $v$, w.r.t. the non re-scaled variable), in which case it approaches zero fast enough for all the moments of the trait distribution to be finite. The particular shapes of the functions $r$, $\alpha$ and $\mathcal{F}$ depend on the details of the limiting processes. 
But, around the {\em critical point} where $\tau=0$ and $h=0$, the functions $r$ and $\alpha$ exhibit {\em universal} behavior that is to a large extent independent of the details of the limiting processes, and implies robust scaling laws.
In particular, we will show that $\alpha(\tau)$, which is a monotonously growing function, jumps from $1$ to $2$ at $\tau=0$, for model class (\ref{eq:SI_ModelA}).
This jump leads to a drastic change in scaling of the moments $\langle x^k\rangle$ in the limit $h\rightarrow 0$. 
Furthermore, we will show for the same model class that, in the limit $h\rightarrow 0$, the population growth rate $r(\tau,0)$, which is also a monotonously growing function,
continuously transitions from the division rate $\omega_2$ (for $\tau>0$ i.e. $\omega_2>\omega_1$), to the growth rate $\omega_1$ (for $\tau \lesssim 0$).
The point $\tau=0$ thus delineates two phases, which we call the {\em growth-dominated phase} (GDP, $\tau<0$) and the {\em division-dominated phase} (DDP, $\tau>0$), respectively. The transition will turn out to be {\em second-order} w.r.t. the order parameter $\langle x\rangle$. 
Let us first derive the steady-state exponential growth rate
$r$. Integrating Eq.~(\ref{eq:HallWake}) w.r.t. $x$, we find that
\begin{eqnarray}\label{eq:SI_omega}
r=\langle b(x)\rangle\,.
\end{eqnarray}
Multiplying Eq.~(\ref{eq:HallWake}) first with $x$ before integration, we see that the $b$-terms drop out and we are left with $(d/dt)(\langle x\rangle_t N(t))=\langle g\rangle_t N(t)$, where $N$ is the particle number. At steady exponential growth, $\dot N/N=r$, and thus, using Eq.  (\ref{eq:SI_omega}), we find that  
\begin{equation}\label{eq:SI_generalDynamics}
\frac{d}{dt}\langle x\rangle_t
=
\langle g(x)\rangle_t
-
\langle x\rangle_t\langle b(x)\rangle_t
\,.
\end{equation}
Plugging Eqs.~(\ref{eq:SI_ModelA}) into Eq.~(\ref{eq:SI_generalDynamics}) yields, for the equilibrium distribution, 
\begin{eqnarray}\label{eq:SI_ModelAstateeq}
\langle x\tilde g(h/x)\rangle=\langle x\rangle (\tau +\langle\tilde b(x)\rangle)\,.
\end{eqnarray}
Let us first consider the GDP ($\tau<0$).
Due to the assumptions made above about $\tilde g$, the l.h.s.~of Eq.~(\ref{eq:SI_ModelAstateeq}) scales like $h$, for small $h$. Hence, in the limit $h\rightarrow 0$, $\langle\tilde b(x)\rangle\rightarrow|\tau|$ and $\langle x\rangle \sim h^0$, since the case $\langle x\rangle \rightarrow 0$ would entail $\langle\tilde b(x)\rangle\rightarrow 0$ and thus contradict positivity of the l.h.s.~of Eq.~(\ref{eq:SI_ModelAstateeq}).
To derive $\alpha$ and reveal the moment scaling, we plug the pure power law ansatz $\psi(x,t)=e^{rt}x^{-\alpha}$ into Eq.~(\ref{eq:HallWake}), and let $u\rightarrow 0$ and $v\rightarrow\infty$ (i.e.~$h \rightarrow 0$), yielding
\begin{equation}\label{eq:SI_alphaeq}
    r+\omega_1(1-\alpha)+\omega_2(1-2^{2-\alpha})=0\,.
\end{equation}
Because $\langle\tilde b(x)\rangle\rightarrow -\tau$, when $h\rightarrow 0$, Eqs.~(\ref{eq:SI_ModelA}) and (\ref{eq:SI_omega}) imply that in the GDP the population growth rate is given by
\begin{equation}\label{eq:rGDP}
    r(\tau,0)=1+\mathcal{O}(\tau^2)\,,\quad\tau <0\,.
\end{equation}{}
In terms of the original parameters this means that the population growth rate is limited by the cellular growth rate, $r\approx\omega_1$, which is larger than the division rate $\omega_2$. This is because most cells will be near the upper limit given by $v$.
Plugging Eq.~(\ref{eq:rGDP}) into Eq.~(\ref{eq:SI_alphaeq}) and keeping only terms up to linear order in $\tau$, we derive
\begin{equation}\label{eq:SI_alphaADP}
\alpha-2=(1+\tau)(1-2^{-(\alpha-2)})\,,
\end{equation}
which has two solutions for $\tau \lesssim 0$ , namely $\alpha=2$, and $\alpha=1+ \tau/(2\ln 2-1)+\mathcal{O}(\tau^2) \lesssim 1\,$. The solution $\alpha=2$ leads to the contradiction $\langle x\rangle \rightarrow 0$. Hence, when we send first $h\rightarrow 0$ and then $|\tau|\rightarrow 0$, we find that, for $k\geq 1$, 
\begin{equation}\label{eq:SI_momentScaling}
\langle x^k\rangle
=
\frac{\int_0^\infty x^{k-\alpha}\mathcal{F}(h/x,x)}{\int_0^\infty x^{-\alpha}\mathcal{F}(h/x,x)}
\sim |\tau|\,,
\end{equation}
because of the pole appearing in the denominator when $\tau$ approaches zero from below and thus $\alpha$ approaches one from below.

In the DDP ($\tau>0$), on the other hand, we derive from Eq.~(\ref{eq:SI_ModelAstateeq}) that 
\begin{equation}\label{eq:DDPscaling}
    \langle x\rangle \sim h/\tau\,,
    \quad\text{for}\quad h\ll\tau\,.
\end{equation}{}
From Eqs.~(\ref{eq:SI_omega}) and (\ref{eq:SI_ModelA}) we further derive the effective population growth rate
\begin{equation}\label{rDDP}
    r(\tau,0)=1+\tau\,,\quad \tau>0\,.
\end{equation}{}
This means that, in this phase, the population growth rate is given by the division rate $\omega_2$, because most cells will be near the lower limit $u$. 
Plugging Eq.~(\ref{rDDP}) into Eq.~(\ref{eq:SI_alphaeq}) we find that $\alpha$ satisfies
\begin{equation}\label{eq:SI_alphaDDP}
\alpha-1=2(1+\tau)(1-2^{-(\alpha-1)})\,.
\end{equation}
This equation has two solutions, $\alpha=1$, and $\alpha \gtrsim 2$, the first solution being incompatible with $\langle x\rangle =\mathcal{O}(h)$. Thus, in the DDP, $\alpha>2$, and Eq. (\ref{eq:FSS}) implies {\em anomalous scaling}, for the positive moments:
\begin{equation}\label{eq:SI_momentScalingDDP}
\langle x^k\rangle\sim
\left\lbrace
\begin{array}{ll}
h^k\,,\quad &k<\alpha-1\\
h^{\alpha-1}\,,\quad &k\geq \alpha-1\,.
\end{array}
\right.
\end{equation}
The universal scaling laws for $\langle x\rangle$ that are expressed in Eqs. (\ref{eq:SI_momentScaling}) and (\ref{eq:DDPscaling}), can be written in the form 
\begin{equation}\label{eq:SI_scaling}
\langle x \rangle 
=
|\tau| f_{\pm}\left(\frac{h}{|\tau|^2}\right)\,,
\quad
h\ll |\tau|\ll 1\,,
\end{equation}
where the two scaling functions $f_{\pm}$, valid for positive and negative $\tau$, respectively, are analytic functions, for small arguments, and both $f_+(y)/y$ as well as $f_-(y)$ converge to positive constants when $y\rightarrow 0$.
This transition is therefore indeed a {\em second order phase transition}: There is a continuous change of {\em order parameter} $\langle x\rangle$, from zero to positive values, when the {\em control parameter} $\tau$ transitions from positive to negative values (for $h=0$), and a universal scaling regime around the {\em critical point} marked by $\tau=h=0$.

For the simplest member of our model class, where $\tilde g(y)=y$ and $\tilde b(y)=y$, $\langle x\rangle$ can be calculated exactly by integrating Eq. (\ref{eq:HallWake}) twice, once with and once without prior multiplication with $x$. We find that
\begin{alignat}{2}\label{eq:SI_modelAorderpar}
\langle x\rangle
=
\frac{-\tau+\sqrt{\tau^2+4h(\tau+1)}}{2(\tau+1)}\,,
\end{alignat}
which immediately verifies the scaling laws derived from Eq.~(\ref{eq:SI_scaling}).

The dynamics of fluctuations around the stationary solution emerge as follows. Setting $\langle x\rangle_t=\langle x\rangle+\epsilon(t)$ and using Eq.~(\ref{eq:SI_generalDynamics}), we find, for $h\ll|\tau|$, that
\begin{equation}\label{eq:SI_slowingdown}
\dot\epsilon(t)
=
-|\tau|\epsilon(t)
+\mathcal{O}(|\tau|)
+\mathcal{O}(\epsilon(t)^2)\,.
\end{equation}
This implies that perturbations decay exponentially on a time scale $|\tau|^{-1}$ (or $|\omega_1-\omega_2|^{-1}$). At criticality, where this time scale diverges, the return to equilibrium follows the much slower algebraic decay $\epsilon(t) \sim t^{-1}$ ({\em critical slowing down}).

In the context of our experimental phytoplankton data, the analyzed phase transition is able to explain $(i)$ the change of the effective population growth rate $r$ (Fig.~\ref{fig:timeDev} (a)), accompanied by $(ii)$ the decline of the mean trait value $\langle x\rangle$ (Fig.~\ref{fig:timeDev} (b)) and $(iii)$ the switch to anomalous scaling of the higher moments $\langle x^k\rangle$ (Fig.~\ref{fig:momentScalingPlusRatios}). 
We hypothesize that, in our experiment, the chlorophyll accumulation rate $\omega_1$ drops below $\omega_2$ due to nutrient depletion (i.e.~$\tau$ crosses zero) and the cultures transition from the GDP to the DDP. Consequently, the effective population growth rate $r$ transitions from tracking $\omega_1$ to tracking $\omega_2$ as described by Eqs.~(\ref{eq:rGDP}) and (\ref{rDDP}). Numerical simulations indicate that an increased death rate $\omega_3$ is required in the DDP to explain the observed decrease in $r$ (see Appendix).
The transition from $\langle x\rangle\sim h^0$ to $\langle x\rangle\sim h^1$ implies that, w.r.t. the non re-scaled variable, $\langle x\rangle$ transitions from scaling with $v$ to scaling with the much smaller $u$. To explain the scaling of the higher moments, we assume scaling relationships $u \sim \ell^{\mu}$ and $v\sim \ell^{\nu}$, where $\ell$ is related to the relevant resource, in our case the light intensity. 
This reflects the intuition that the light availability determines the `optimal' chlorophyll content in each phase. 
Note that, as $\ell$ drops out of our final scaling results, we do not have to specify how to measure the resource. We only need to assume that, as the resource varies, $u$ and $v$ are related by a scaling relationship. 
In the GDP, where $\alpha < 1$, we derive from $\langle x^k\rangle \sim h^0$ that, w.r.t. the non re-scaled variable, $\langle x^k\rangle\sim v^k$, for all $k\geq 1$. This results in trivial moment scaling $\langle x^k \rangle \sim \langle x \rangle^k$ and the observed approximate data collapse. The deviation from the collapse at very low trait values is attributed to variations of $u$.
In the DDP, where $\alpha \gtrsim 2$, we derive from (\ref{eq:SI_momentScalingDDP}) that $\langle x \rangle \sim u \sim \ell^{\mu}$, while 
\begin{equation}\label{eq:momentScalingReal}
\langle x^k \rangle \sim \ell ^{\nu k + (\alpha-1)(\mu-\nu)} 
\sim \langle x \rangle^ {(\nu/\mu) k + (\alpha-1)(1-\nu/\mu)}\,,
\end{equation}
for $k>1$. This becomes, for $\alpha \gtrsim 2$, approximately $\langle x^k \rangle \sim \langle x \rangle^ {1+ (\nu/\mu) (k -1)}$, which results in the observed anomalous moment scaling with $\gamma=\nu/\mu$ (Fig. \ref{fig:momentScalingPlusRatios} (d)).
Additionally, our theoretical results are compatible with the data collapse and trivial moment scaling observed in Ref.~\cite{giometto2013scaling}, which in terms of our model class would mean that the system remained in a GDP, perhaps due to an abundance of nutrients. 

Obtaining $\alpha$ from the experimental data through moment scaling was possible because $\tau$ and $h$ could be modified independently: While $\tau$ appears to be primarily influenced by the nutrients, $h$ appears to vary primarily with the light condition (or perhaps the species in case of size distributions, e.g. \citep{giometto2013scaling}).
This allowed us to study the scaling of $\langle x\rangle$ with respect to $h$, but our model also predicts scaling laws with respect to $\tau$.
From Eq.~(\ref{eq:SI_scaling}) we derive two scaling laws, respectively, for the mean $\langle x\rangle$ and its {\em susceptibility} to changes of external conditions (measured in terms of $h$),
\begin{equation}\label{eq:tauscaling}
    \langle x\rangle|_{{h=0}} \sim |\tau|\,,\quad \text{for}\quad\tau<0\,;\quad
    \frac{\partial \langle x\rangle}{\partial h}\Bigr|_{h=0} \sim |\tau|^{-1}\,.
\end{equation}
In addition, according to Eq.~(\ref{eq:SI_slowingdown}), the time, $T$, it takes to return to an equilibrium trait distribution after a perturbation diverges according to the power law
\begin{equation}\label{eq:RTE}
    T\sim |\tau|^{-1}\,.
\end{equation}
Thus $\tau$ can be used to relate the temporal and trait scales.
Moreover, the scaling laws in Eqs.~(\ref{eq:tauscaling}) and (\ref{eq:RTE}) can be combined to allow model validation without requiring measuring $\tau$ itself. For example the response of trait distributions to external perturbations could be measured in an array of chemostats kept at different distances to the critical point (e.g.~by a careful tuning of nutrient content).

The reader might wonder about caveats in our approach.
First, our analytic results are derived for equilibrium trait distributions, which we likely do not have in our experimental data. However, simulations indicate that transient distributions follow the same scaling behaviour (see Appendix). Second, outlier discarding may substantially affect the moments of power-law distributions and therefore requires careful justification. A large fraction of flow-cytometry pulses from cells classified as outliers show multimodality, indicating preparation for cell division (division is instantaneous in our model). Additionally, outliers may be the result of errors in the scanning flow cytometer (cells becoming stuck and moving below the expected rate of flow, or data processing errors), resulting in anomalously large readings. This small number of outlying cells can dominate the higher distribution moments and lead to substantial fluctuations in those moments. While we believe that on biological grounds our outlier rejection is well-founded, such a step could in principle alter the moment scaling and force us to enlarge the class of models that could explain the observed scaling laws (see Appendix).  
Third, our model class is based on single traits, but it is known that this may not suffice to define cell division rates (e.g. the cell {\em age} can play an important role \cite{KennardEColiDivision}), which indicates a promising avenue for future generalizations of our work. Fourth, we have focused on exponentially growing populations, and neglected interactions that are salient in natural communities. However, when including a simple type of mean-field interaction that modifies the probability of cell death (e.g. through a resource-competition death rate that is proportional to some extensive quantity calculated over the whole population), we recover the same critical point and phase transition. 

With rare exceptions \citep{Merico2014}, scaling that constrains or defines relationships between traits and the environment, trade-offs, and growth and reproduction of organisms (e.g. \citep{maranon_2015_CellSizePhytoMetabolism, litchman2008trait, Barton2013, Smith2009, Merico2009}), is assumed to be static. Our approach shows quantitatively how some of these scaling relationships might change under changing environmental conditions, and may thereby prove useful in augmenting these existing trait-based modelling approaches. 
While we have focused here on chlorophyll-a and light intensity, in future investigations other pairs of trait and environment variables (such as size and temperature) might be related through our scaling laws as well.

\section*{Acknowledgements}
The Authors gratefully acknowledge continued in-depth discussions with Amos Maritan and Andrea Rinaldo.
This work was supported by the Swiss National Science Foundation SNF Grant Nr. 159660, and the Eawag Discretionary Fund.

\section{Appendix}

\subsection{Numerical experiments}
\label{sec:SI_numericalExperiments}

In order to corroborate our theoretical results and check to what extent they generalize to the transients, we adapted Gillespie's algorithm (see below) to simulate the simplest representative from our model class, where growth, division, and death rates are given by, respectively, $g(x) = \omega_1(u+x)$, $b(x) = \omega_2(1+x/v)$, and $d(x) = \omega_3$. 
The base rates $\omega_1$, $\omega_2$ and $\omega_3$ were kept fixed during the first, fast growth phase, then changed abruptly once a certain population size was reached and then kept again fixed during the second, slow growth phase. Their numerical values were chosen so as to reproduce, approximately, the observed population growth rates as well as the observed rate of change of the first moment (Fig. \ref{fig:timeDev}). Furthermore, $\omega_1$ and $\omega_2$ were chosen such that $\tau$ remained close to the critical point (i.e. slightly negative in the first phase and slightly positive in the second). An increased death rate $\omega_3$ was needed in the second phase in order to explain the observed decline in population growth rate at the transition point. 
The values for $u$ and $v$, for the simulations corresponding to each light setting, were kept fixed during each numerical experiment, such that: i) they follow the scaling relationships with an arbitrary light resource $l$ as described above Eq. (\ref{eq:momentScalingReal}); and ii) $\gamma=\nu/\mu$ was consistent with the experimental moment scaling.
All the parameter values are summarized in Table \ref{tab:transientParameters}.

\begin{figure}[h]
\centering

      \includegraphics[width=\columnwidth, trim=0 0 0 0, clip]{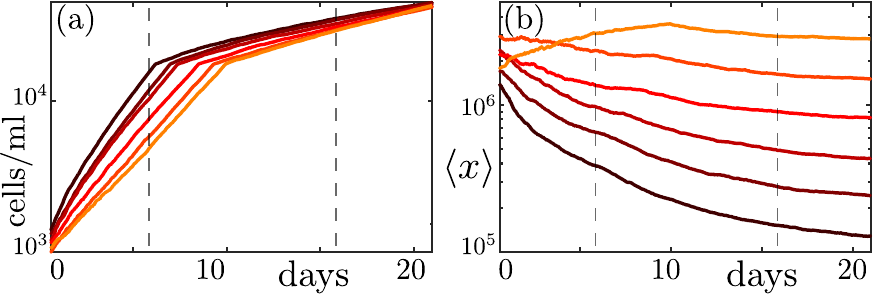}
   
\caption{Simulation results for six cultures (different colors), where the phase transition (i.e.~the parameter switch) is initiated when a certain population size is reached. Dynamics of cell-density (a) and the mean chlorophyll content (b). Days 6 and 16 indicated by dashed lines (cf. Fig. \ref{fig:transientsMomentScaling}).}
          
\label{fig:transientsTimeDev}
\end{figure}

\begin{figure}[h]
\centering

      \includegraphics[width=.9\columnwidth, trim=0 0 0 0, clip]{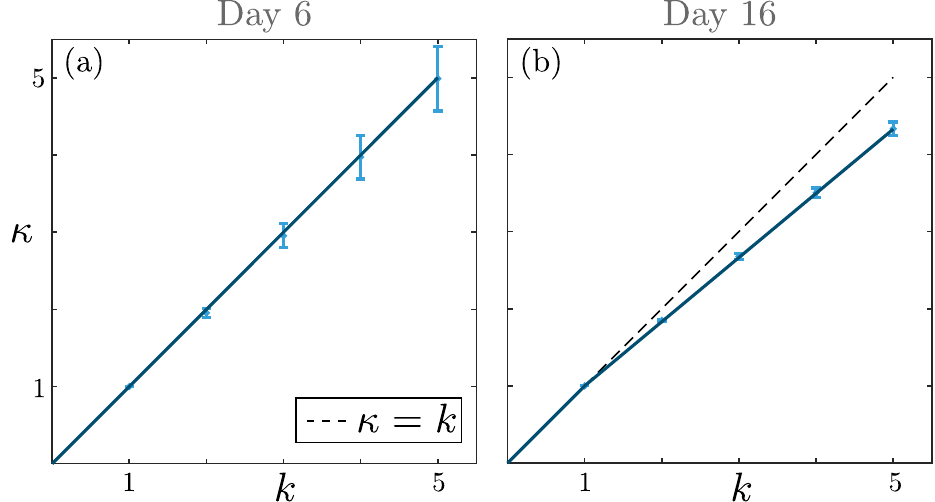}
   
\caption{Moment scaling $\langle x^k\rangle \sim\langle x\rangle ^{\kappa(k)}$ from the simulations at the time-points indicated in Fig. \ref{fig:transientsTimeDev}, before (a) and after (b) the transition.}
          
\label{fig:transientsMomentScaling}
\end{figure}

While our simulations did not and were not intended to reproduce all the features seen in our data, they do approximately reproduce $(i)$ the observed decrease of the population growth rate (Fig. \ref{fig:transientsTimeDev} (a)), $(ii)$ the observed decrease of the mean chlorophyll content (Fig. \ref{fig:transientsTimeDev} (b)) as well as $(iii)$ the observed transition from trivial to anomalous moment scaling (Fig. \ref{fig:transientsMomentScaling}). In particular, the anomalous moment scaling became apparent almost immediately after the transition, and long before an equilibrium trait distribution was reached. 

\begin{table}
\centering

\begin{tabular}{| l | l | l | l | l |  l |}
  \hline                 
  Phase & $\omega_1$ & $\omega_2$  & $\tau$             & $\alpha$  & $\omega_3$ \\ \hline      
  GDP (all cultures) & $0.92$ & $0.9$ & $-0.02$ & $\approx 0.95$ & $0.62$ \\
  DDP (all cultures) & $0.86$ & $0.9$ & $0.05$ & $\approx 2.16$ & $0.9$ \\
  \hline  
\end{tabular}

\vspace{0.5cm}

\begin{tabular}{| l | l | l |}
  \hline                 
Culture & $u$ & $v$ \\ \hline   
1  &  $0.11 \cdot 10^5$ & $0.22 \cdot 10^7$ \\
2 &  $0.23  \cdot 10^5$ & $0.37 \cdot 10^7$ \\
3 &  $0.48  \cdot 10^5$ & $0.61 \cdot 10^7$ \\
4 &  $0.99  \cdot 10^5$ & $1.01 \cdot 10^7$ \\
5 &  $2.04  \cdot 10^5$ & $1.68 \cdot 10^7$ \\
6  &  $4.21  \cdot 10^5$ & $2.80 \cdot 10^7$ \\
  \hline  
\end{tabular}
\caption{Parameter values in model simulations.}
\label{tab:transientParameters}
\end{table}

\subsection{Generalised Gillespie algorithm}
\label{sec:SI_Gillespie}

The Gillespie algorithm \citep{gillespie1976} offers an efficient way of simulating large numbers of particles with constant reaction rates. Rather than simulating forward in time with constant time increments, the distribution of reaction event times is computed, and event times are then drawn from this distribution.

In our models, the only `reactions' are cell division and cell death events. However, the associated rates $b(x(t))$ and $d(x(t))$ may depend on the state of the cells, which vary in time due to growth of the cells. 
For the simple model we used in our numerical simulations, the death rate is constant in time and the same for all cells: $d=\omega_3$. Thus, the minimal survival time for $N$ cells is distributed exponentially with decay rate $N\omega_3$.
The division rate, due to its dependence on $x(t)$, is different for each cell and varies over time. Thus, the time when the next division event occurs (if not preceded by a death event) needs to be drawn individually, for each cell. For our simple model, the corresponding "survival function", $s(t)$, which is the probability for the cell not to divide until time $t$, can be calculated analytically.
With $g(x) = \omega_1(u+x)$, a cell with initial value $x_0$ grows according to $x(t) = (x_0+u) e^{\omega_1 t} -u$. Plugging this dependence into $b(x) = \omega_2(1+ x/v)$, we solve $\dot s(t)=-b(x(t))s(t)$ and find that
\begin{multline}
\log s(t) = \frac{\omega_2}{\omega_1}\frac{x_0+u}{ v} + \omega_2 \left(\frac{u}{v}-1\right) t \\
- \frac{\omega_2}{\omega_1}\frac{x_0+u}{ v} e^{\omega_1 t} \,.
\label{eq:SI_survivalProb}
\end{multline}
To draw a division time, $t_d$, from the associated distribution, we draw a random value $k$ from a uniform distribution in the interval $[0,1]$ (i.e., `draw a survival probability') and solve $\log s(t_b) = \log k$ for $t_b$. 
In systems with many cells the division times will generally be very small, so we expand Eq.~(\ref{eq:SI_survivalProb}) up to second order in $t$ and solve the resulting equation, from which we derive the approximate division time 
\begin{multline}
t_b = -\frac{1}{\omega_1}\frac{x_0+v}{x_0+u} \\
+ \frac{1}{\omega_1}\frac{v}{x_0+u}\sqrt{  \left(\frac{x_0}{v}+1\right)^2 - 2\frac{\omega_1}{\omega_2}\frac{x_0+u}{v} \log k }  \,.
\end{multline}

After a division time has been simulated, individually for each cell, and the shortest time until the next death event occurs has been simulated, for all cells simultaneously, the smallest of these times is chosen and the corresponding event is executed. 

\subsection{Removal of Outliers}
\label{sec:Outliers}

For the computation of the moment scaling shown in Figures \ref{fig:momentScalingPlusRatios} and \ref{fig:momentScalingAllCultures}, outliers (between $\approx 1\%$ and $\approx 5\%$ cells with largest chlorophyll content) were removed from the data to reduce the noise and reveal the moment scaling. The outliers are classified as the data points that are above the quantile $q=(q_u-q_l) \cdot 2+q_u$, where $q_l$ is the $25\%$ quantile and $q_u$ is the $75\%$ quantile. The percentage of outliers is typically larger in the GDP than in the DDP.
These outliers dominate particularly the higher moments and obscure the moment scaling completely. They are thought to originate mainly from cells in the process of cell division at the time of measurement, as well as faulty measurements. 
Indeed, a large fraction of the cells classified as outliers show evidence of being in the process of dividing. Between $\approx 30\%$ and $\approx 50\%$ of the flow-cytometry pulses (recording the forward- and sideward scattered light and chlorophyll fluorescence signals through time) within the outliers show multimodality, suggesting that the cell content is being split up in preparation for division. 
Among the cells below the threshold (above which they are classified as outliers), only between $\approx 5\%$ and $\approx 20\%$ of the flow-cytometry pulses show multimodality.

However, such a removal of outliers could potentially also change the moment scaling. If a constant fraction of outliers above a quantile $q_u$ was removed in our model, the moment scaling would also in the DDP revert to a trivial moment scaling. This is because any upper quantile will eventually scale as $q_u\sim u$, dominating the scaling of all higher moments. However, our numerical experiments show that also in this case anomalous moment scaling is observed during a transient, while the mean already scales like $u$ but the upper quantile still scales like $v$.
Such a spurious appearance of anomalous moment scaling during a transient could even emerge if $\alpha$ crosses $1$ in a continuous rather than discontinuous manner.
A more general type of phase transition, including such continuous changes of $\alpha$, could be defined using a different order parameter (e.g.~the fraction of cells with trait $x>v$), which then comes with a larger universality class of limiting mechanisms (including, for instance, a hard cut-off preventing cells with trait values $x<u$ from dividing). Although we believe our outlier rejection to be well justified, this more general phase transition might also be consistent with our data. It would therefore be valuable to focus future experimental work not only on scaling, but also on the class of limiting processes required to capture growth and division dynamics in unicellular species of interest.

\subsection{Additional Experimental Evidence}
\label{sec:SI_experimentEvidence}

Here we show evidence that the phase transition occurs not only in the directly illuminated $Pseudokirchneriella$ cultures that were shown in the main text, but also in the $Microcystis$ cultures and in shaded cultures of both species, which received only light filtered by the competitor (see Ref. \citep{fontana_2019_lightExp}).

\begin{figure}[h]
\centering

\includegraphics[width=1\columnwidth, trim=0 0 0 0, clip]{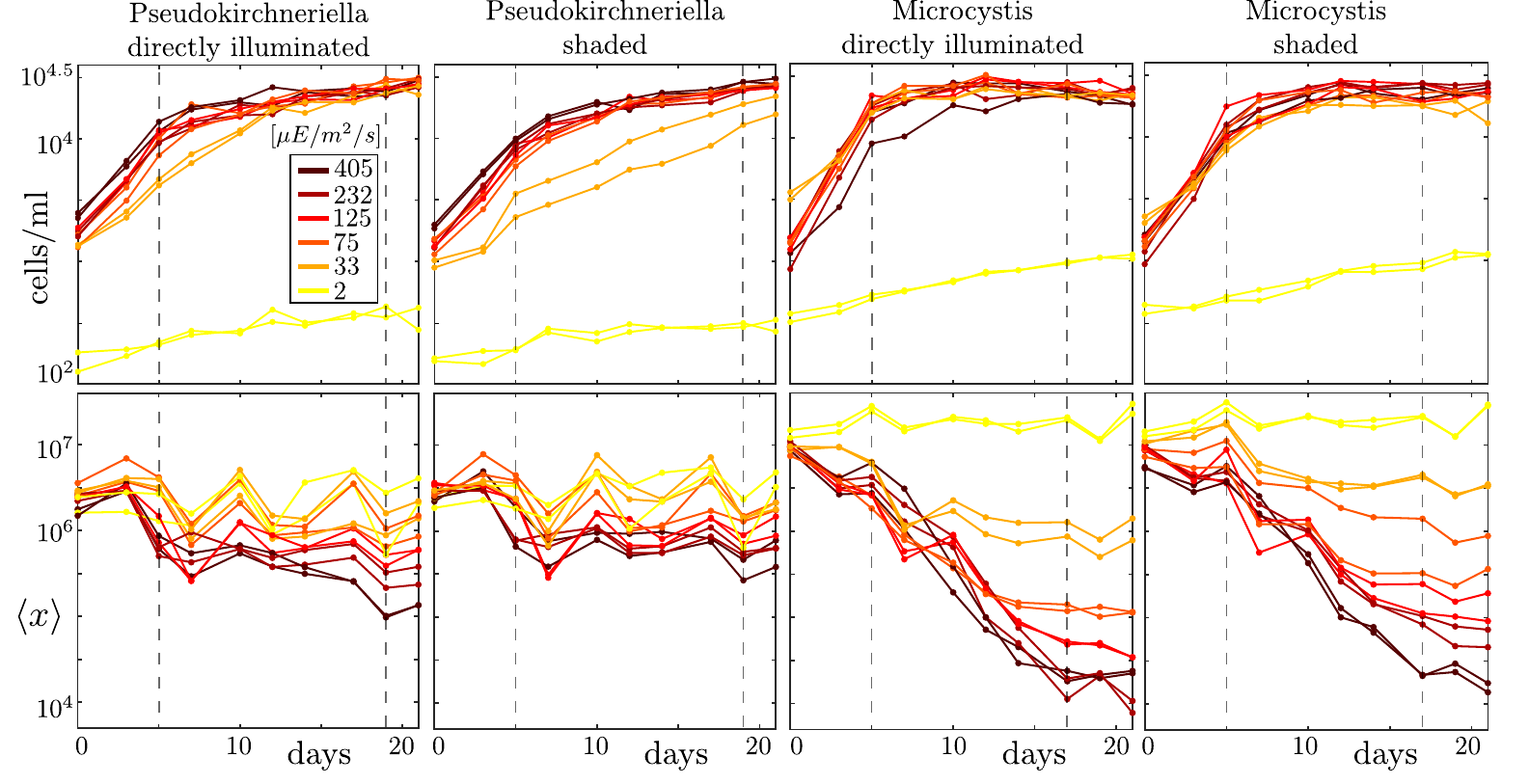}
   
\caption{
Dynamics of the cell-density (top panel) and the mean cellular chlorophyll-a content (bottom panel), for different light settings (photon irradiance, indicated by color) and all experimental conditions. Dashed lines indicate time points (days 5 and 19 in Pseudokirchneriella, days 5 and 17 in Microcystis) shown in Figs.~\ref{fig:collPseudoShaded}, \ref{fig:collMicro}, \ref{fig:momentScalingAllCultures}. }
          
\label{fig:timeDevAllCultures}
\end{figure}

\begin{figure}[h]
\centering

\includegraphics[width=\columnwidth, trim=0 0 0 0, clip]{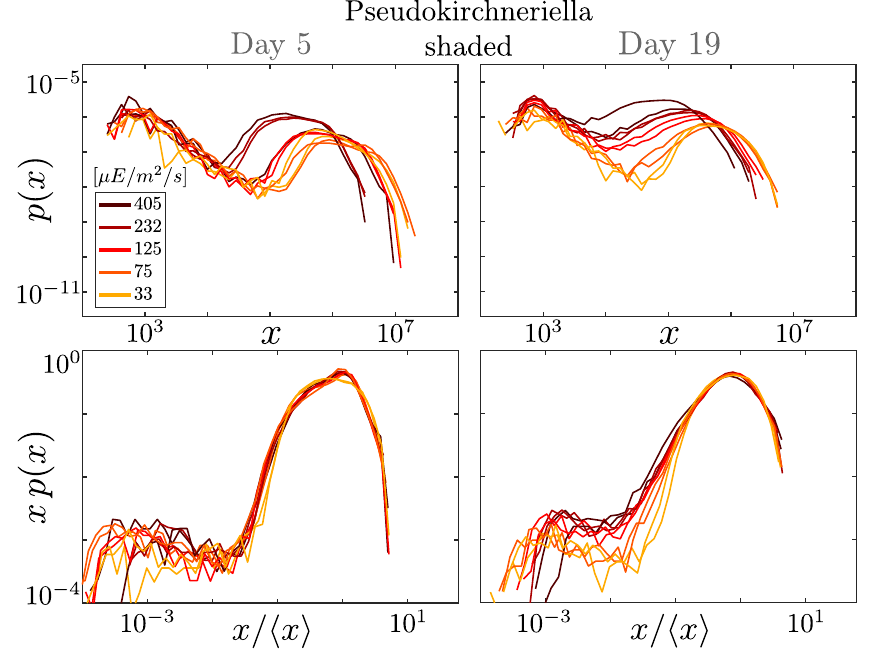}
   
\caption{
Chlorophyll-a distributions (top) of shaded Pseudokirchneriella cultures and their rescaled versions (bottom) at days 5 and 19 from the GDP (left) and DDP (right) respectively. The GDP data collapse (bottom left) breaks down in the DDP (bottom right).
}
          
\label{fig:collPseudoShaded}
\end{figure}

\begin{figure}[h]
\centering

\includegraphics[width=\columnwidth, trim=0 0 0 0, clip]{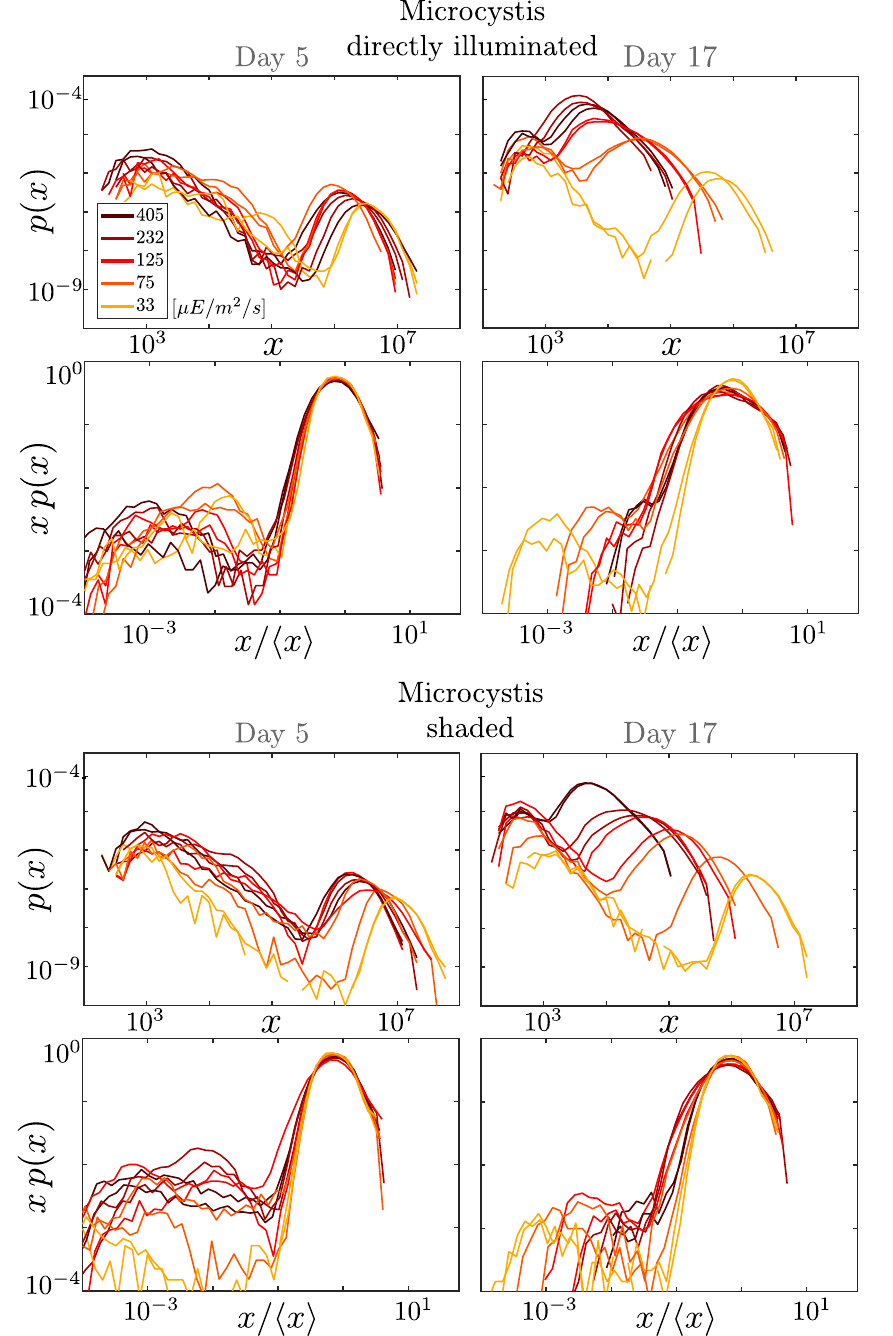}
   
\caption{
Chlorophyll-a distributions of Microcystis cultures (top panels) and their rescaled versions (bottom panels) at days 5 (left) and 17 (right). The approximate GDP data collapse (left) breaks down in the DDP (right).
}
          
\label{fig:collMicro}
\end{figure}

\begin{figure}[h]
\centering

\includegraphics[width=\columnwidth, trim=0 0 0 0, clip]{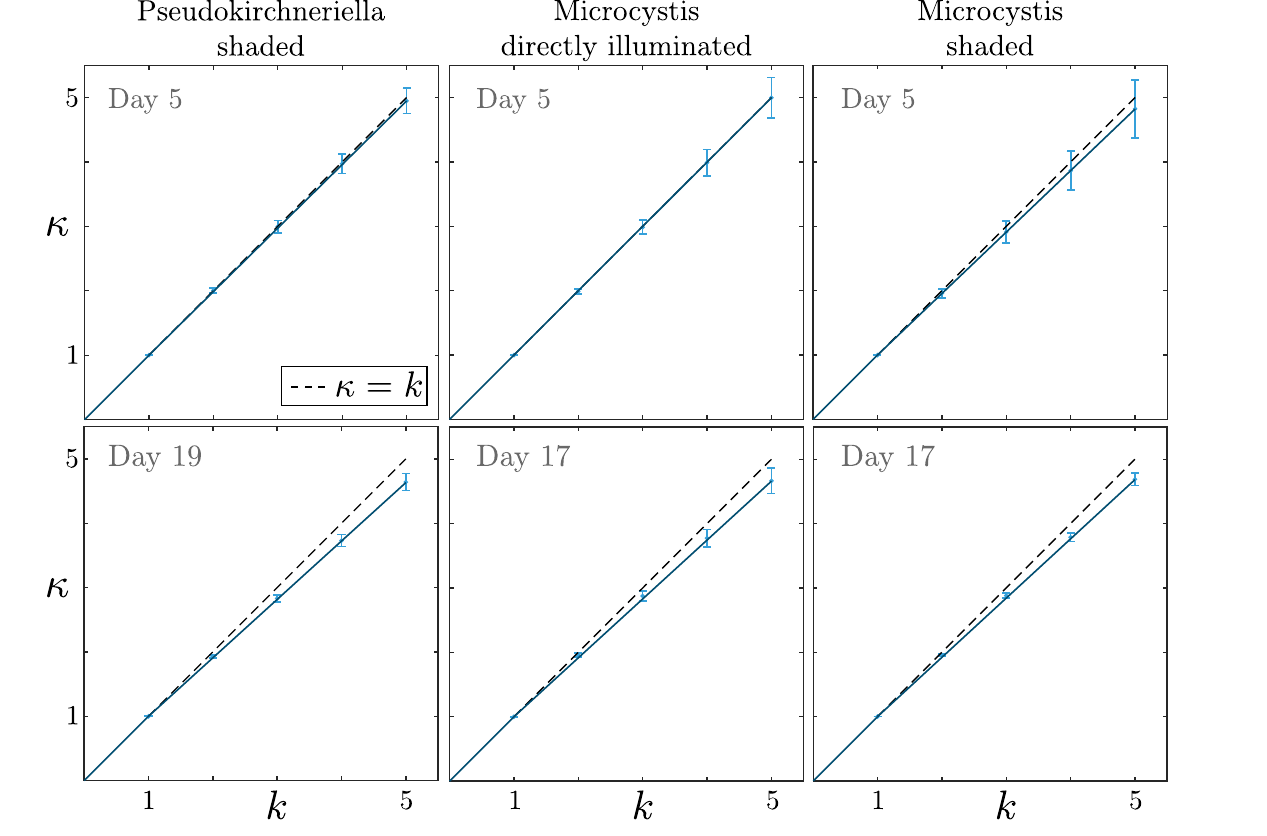}
   
\caption{Moment scaling $\langle x^k\rangle\sim\langle x\rangle^{\kappa(k)}$ in Pseudokirchneriella (shaded) and Microcystis (directly illuminated and shaded) cultures in the GDP (upper panels) and the DDP (lower panels).}
          
\label{fig:momentScalingAllCultures}
\end{figure}

Initially, all cultures show behaviour consistent with a GDP, where we observe fast exponential population growth (Fig. \ref{fig:timeDevAllCultures}, top panel), data collapse (Figs. \ref{fig:collPseudoShaded}, \ref{fig:collMicro}), and trivial moment scaling of the form $\langle x^k\rangle \sim \langle x\rangle ^k$, for all observable positive moments (Fig. \ref{fig:momentScalingAllCultures}, top panel).
Later in the experiment, all cultures, except those in the lowest light intensity, appear to transition to the DDP, where the population growth rate decreases, there is no data collapse, and the anomalous moment scaling $\langle x^k\rangle \sim \langle x\rangle ^{1+\gamma (k-1)}$, with $\gamma<1$, is observed (Fig. \ref{fig:momentScalingAllCultures}, bottom panel). 
In all cultures, except those in the lowest light intensity, we observe a strong decrease in mean chlorophyll content throughout the experiment (Fig. \ref{fig:timeDevAllCultures}, bottom panel).
In Figures \ref{fig:collPseudoShaded}, \ref{fig:collMicro}, and \ref{fig:momentScalingAllCultures} different time points were selected in Pseudokirchneriella and Microcystis (day 19 and 17, respectively) to represent the  moment scaling and distributions in the DDP. This is because Microsystis became increasingly volatile towards the end of the experiment, at which the cultures collapsed, thereby strongly increasing the uncertainty in the estimation of the higher moments. We thus selected a slightly earlier date for these cultures.
It is evident that the features characteristic for our phase transition (change in population growth rate and moment scaling) can already be observed before the systems reach an equilibrium trait distribution.

Here, we show all data including also the lowest light intensity cultures that were not shown in the main text. The cultures from this light intensity all show very slow population growth as well as little change in the mean chlorophyll-a content over time (see Fig. \ref{fig:timeDevAllCultures}). From population and trait dynamics it is therefore unclear whether these cultures underwent the phase transition, which is why they are omitted from the data collapse and the computation of the moment scaling.

The uncertainty in the moment scaling in the GDP remains relatively large in some cultures even after the removal of outliers. This  may be due to the fact that the cells are still adapting to the experimental environment (individual cells can influence the higher moments strongly, and close to the critical point, convergence may take particularly long; see Eq.~(\ref{eq:SI_slowingdown})). 

\bibliography{refs}

\end{document}